\documentclass[referee]{raa} 

\usepackage{graphicx,times}             
\usepackage{natbib}
\usepackage{amssymb,amsmath}
\usepackage{mathrsfs}         
\usepackage{color}

\newcommand{\feii}{\ifmmode {\rm Fe\ II} \else Fe~{\sc ii}\fi}

\bibpunct{(}{)}{;}{a}{}{,}

\usepackage[pagebackref=true]{hyperref}

\begin{document}

  \title{Correcting the Contamination of Second-order Spectra: Improving H$\alpha$ Measurements in Reverberation Mapping Campaigns }

   \volnopage{Vol.0 (20xx) No.0, 000--000}      
   \setcounter{page}{1}          

   \author{Wen-Zhe Xi 
      \inst{1,2,3}
   \and Kai-Xing Lu
      \inst{1,3}
   \and Hai-Cheng Feng
      \inst{1,3}
    \and Sha-Sha Li
      \inst{1,3}
    \and Jin-Ming Bai
      \inst{1,3}
    \and Rui-Lei Zhou
      \inst{2,4}
    \and Hong-Tao Liu
      \inst{1,3}
    \and Jian-Guo Wang
      \inst{1,3}
   }

   \institute{Yunnan Observatories, Chinese Academy of Sciences,
             Kunming 650011, China; {\it hcfeng@ynao.ac.cn}\\
        \and
            University of Chinese Academy of Sciences, Beijing 100049, People’s Republic of China; \\
        \and
             Key Laboratory for the Structure and Evolution of Celestial Objects, Chinese Academy of Sciences, Kunming 650011,China; 
             {\it lukx@ynao.ac.cn}\\
        \and
            CAS Key Laboratory of FAST, National FAST, National Astronomical Observatories,  Chinese Academy of Sciences, Beijing 100101, China \\
\vs\no
   {\small Received 20xx month day; accepted 20xx month day}}

\abstract{Long-term spectroscopic monitoring campaigns on active galactic nuclei (AGNs) provide a wealth of information about its interior structure and kinematics. However, a number of the observations suffer from the contamination of second-order spectra (SOS) which will introduce some undesirable uncertainties at red side of spectra. In this paper, we test the effect of SOS and propose a method to correct it in the time domain spectroscopic data using the simultaneously observed comparison stars. Based on the reverberation mapping (RM) data of NGC~5548 in 2019, one of the most intensively monitored AGNs by Lijiang 2.4-meter telescope, we find that the scientific object, comparison star, and spectrophotometric standard star can jointly introduce up to $\sim$30\% SOS for Grism 14. This irregular but smooth SOS significantly affects the flux density and profile of emission line, while having little effect on the light curve. After applying our method to each spectrum, we find that the SOS can be corrected effectively. The deviation between corrected and intrinsic spectra is $\sim$2\%, and the impact of SOS on time lag is very minor. This method makes it possible to obtain the $\rm{H\alpha}$ RM measurements from archival data provided the spectral shape of the AGN under investigation does not have a large change.
\keywords{techniques: spectroscopic --- methods: data analysis --- galaxies: individual: NGC 5548 --- quasars: emission lines}
}

   \authorrunning{Xi et al.}            
   \titlerunning{Correcting the Contamination of SOS}  

   \maketitle
%
%

%
\section{Introduction}           
\label{sect:intro}
Active galactic nuclei (AGNs), one of the most luminous objects in the universe, are powered by gas accretion onto supermassive black holes (SMBHs). Broad emission lines are the most prominent feature of AGNs providing important information to study the physical properties of SMBH and its surroundings. In the context of the unified model \citep{An93}, the broad emission lines originate from the broad line region (BLR) located at thousands gravitational radii away from the central SMBH, and are broadened by the Doppler motions of clouds in BLR under the strong gravitational field of SMBH. According to the photoionization model, broad emission lines are driven by central ionizing source. Analyzing the delayed response [or time lag ($\tau$)] of the broad emission lines to the continuum variations will provide information on the scale and structure of BLR, and the mass of SMBH can be derived by 
\begin{equation}\label{eq1}
    M_{\bullet} = f_{_{\rm{BLR}}}\frac{R_{_{\rm{BLR}}}v^2}{G},
\end{equation}
where $f_{_{\rm{BLR}}}$ is a virial factor determined by the geometry and kinematics of the BLR gas, $R_{_{\rm{BLR}}}$ is the responsivity weighted radius of the BLR which is derived by $R_{_{\rm{BLR}}} = \tau \times c$ ($c$ is the speed of light), and $v$ is a measurement of velocity of clouds, e.g., FWHM ($v_{_{\rm{FWHM}}}$) or the ``line dispersion" ($\rm{\sigma_{line}}$; see \cite{Peterson+etal+2004} and references therein). This way is called reverberation mapping \citep[RM;][]{BM82, Peterson+etal+1993}, more details about the parameters are well introduced by \cite{Du+etal+2014}. 

As the number of RM experiments increased, it is gradually found that $R_{_{\rm{BLR}}}$ usually showed tight correlation with continuum luminosity ($L$) where $R_{_{\rm{BLR}}}$ and $L$ are usually measured from $\rm{H\beta}$ emission line and monochromatic luminosity at 5100 \AA\ (\citealt{Kaspi+etal+2000, Denney+etal+2010, Bentz+etal+2013}), respectively. Consequently, the mass of SMBH can be estimated using single-epoch spectrum which is convenient to large spectral surveys such as the Sloan Digital Sky Survey (SDSS; e.g., \citealt{Shen+etal+2014}). For high quality RM data, it is possible to reconstruct the kinematics and geometry of BLR via velocity-resolved time-lag (\citealt{De18, Fe21a, Li+etal+2022, Vi22}) because the velocity and lag of a cloud are determined by its position in BLR. While RM technique is simple and widely used in the literature, it based on two fundamental assumptions that BLR clouds are photoionized by the central ionizing source and dominated by gravity of the central SMBH \citep{Peterson+etal+1993}, which needs to be certified by observational evidence. The photoionization model also predicts a stratified radial structure of the BLR due to ionization energy and optical depth, indicating that we can compare the lags and velocities of different emission lines to valid the two assumptions \citep[e.g., $\tau \propto v^{2}_{\rm{FWHM}}$ relationship;][]{Fa17}. Multi-line RM is therefore an alternative avenue for the study of ionization and virialization properties in BLR. 

So far, about 100 AGNs have been successfully measured the mass of SMBHs by RM method \citep[e.g.,][]{Be10a, Hu21}, but most of them focused on H$\beta$, leading to a H$\beta$ based $R_{_{\rm{BLR}}} - L$ relationship. This makes it difficult to investigate the multi-line properties. Moreover, H$\beta$ emission line will be failed in some cases such as type 1.8/1.9 AGNs \citep{O81} or infrared spectra of some high redshift sources. H$\alpha$ is the strongest optical emission line at low $z$ and it is less blended with the optical \feii\ resulting in a higher Signal-to-Noise Ratio (S/N) and more reliable profile measurements. In some AGNs with weak broad emission feature, H$\alpha$ is usually the only detectable optical broad line (e.g., type 1.9 AGNs), suggesting that it is important to investigate the properties of H$\alpha$ emitting regions. Unfortunately, less than 20 AGNs have been reported H$\alpha$ RM results \citep[e.g.,][]{Be10a}, and only several of them have velocity-resolved measurements \citep{Be10b, Fe21a}. Thus, it is necessary for new spectroscopic monitoring programs to expand the H$\alpha$ RM sample. It is not easy to achieve this goal in short term because RM observations are very time consuming. Considering the fact that most of previous reverberation experiments focused on the local ($z <$ 0.3) AGNs, it is possible to observe H$\alpha$ in some sources \citep[e.g., SEAMBH project;][]{Du+etal+2014}. We might utilize these data to obtain the H$\alpha$ lags, and even analyze the velocity-resolved lags.

Since previous observation strategies are primarily designed to optimize for H$\beta$, H$\alpha$ might suffer from some observational effects. For example, the telluric absorption at wavelength larger than 6800 \AA, which would significantly contaminate H$\alpha$ emission line, were often ignored. Furthermore, most H$\beta$-based observations usually adopted relatively bluer spectrographs \citep[e.g.,][]{Lu+etal+2019}, which might cause the second-order spectrum (here after SOS) to superimpose around H$\alpha$ wavelength. The telluric absorption can be corrected by theoretical telluric absorption spectra \citep{Ka15} or supplementary observations of telluric standard star, though both methods can be affected by changes in the weather conditions. The contamination of SOS is usually solved via a dichroic filter which can divide red and blue light into separate channels \citep[e.g.,][]{OG82}, or an order-blocking filter which can block the UV/blue photons \citep[e.g.,][]{Feng+etal+2020}. However, previous H$\beta$-based RM campaigns rarely adopted either of the two observation schemes. If we can precisely obtain the atmospheric profiles in real time, both telluric absoption and SOS will be accurately corrected. Although, most observatories do not provide the information, we can still correct H$\alpha$ data in this way when there are simultaneously observed comparison stars. As the object and comparison star are under the same weather conditions (e.g., airmass, seeing, clouds, etc.). In fact, there were indeed some previous RM campaigns that used comparison stars for flux calibration \citep[e.g.,][]{Maoz+etal+1990,Kaspi+etal+2000,Du+etal+2014,Hu21, Fe21b}. Thereby, it is possible to extract H$\alpha$ RM results from archival data. It has been proved that the telluric absorption can be well corrected by a comparison star \citep{Feng+etal+2020,Lu+etal+2021}, while the SOS still requires effort. 

Starting in 2012 October, a large RM project \citep{Du+etal+2014} was carried out on the 2.4 m telescope located at Lijiang, aimed at increasing the RM number of AGNs with high Eddington ratio. The project have monitored $\sim$50 AGNs and most of them also observed H$\alpha$ emission line. The varying degrees of contamination by SOS in these spectral data motivated us to investigate this effect. 

The paper is organized as follows. In Section~\ref{sec2}, we present the correction method of SOS. Application to NGC~5548 is presented in Section~\ref{sec3}. Discussion is in Section~\ref{sec4}, and summary results in Section~\ref{sec5}. 

\section{METHOD}
\label{sec2}

   \begin{figure}
   \centering
   \includegraphics[width=\textwidth, angle=0]{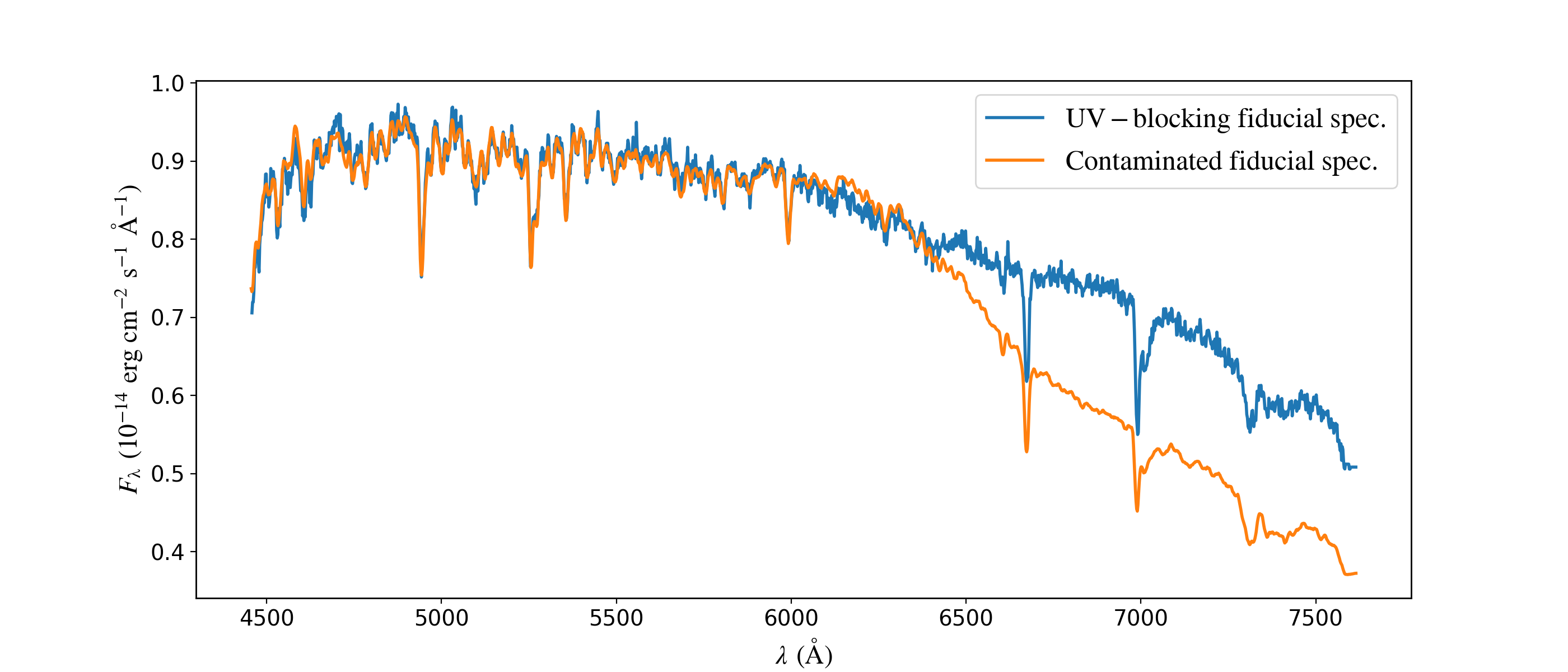}
   \caption{\footnotesize
   Fiducial spectrum of the comparison star for NGC~5548. The orange line is the contaminated spectrum generated from an 81-day observation with contamination starting from 6300$\mathrm{\mathring{A}}$. The blue line is the uncontaminated fiducial spectrum generated from a 7-day observation with a UV-blocking filter.}
   \label{Fig1}
   \end{figure}
   
Flux calibration strategy based on comparison star is commonly used for spectroscopy monitoring projects \citep[e.g.,][]{Maoz+etal+1990, Kaspi+etal+2000, Fe21a}. In some RM campaigns, there are several nearby comparison stars within 1$^{\circ}$ of targets that are simultaneously observed by multi-slit \citep{Wi21} or fibers \citep{Sh16}, while others can only obtain one single comparison star \citep{Du+etal+2014, Lu+etal+2022}. Due to the identical observation conditions of target and comparison star, this approach can achieve high-accuracy relative flux calibration even during poor weather. For most previous H$\beta$-focused spectra, the flux calibration generally consists of the following steps:
\begin{enumerate}

\item The absolute flux calibration of each comparison star was performed using the spectrophotometric standard star.

\item A fiducial spectrum of the comparison star is generated by combining the spectra observed in clear nights.

\item Each comparison spectrum is compared to the fiducial spectrum to derive sensitivity functions which can be directly used to calibrate flux of the target.

\end{enumerate}
However, the above process cannot correct the SOS contributions because standard star, comparison star, and object all introduce different contamination.

In order to correct the contamination of SOS, we start with the calibrated flux of object
\begin{equation}\label{eq2}
    {O}(\lambda) = \frac{{C}_{\rm fid}(\lambda)}{{C}_{\rm cou}(\lambda)}{O}_{\rm cou}(\lambda),
\end{equation}
where ${O}_{\rm cou}$ and ${C}_{\rm cou}$ are the observed spectra (in counts) of object and comparison star, respectively, ${C}_{\rm fid}$ is the fiducial spectrum of comparison star that can be expressed as
\begin{equation}\label{eq3}
    {C}_{\rm fid}(\lambda) = \overline{\frac{10^{\rm E(\lambda) \cdot (\rm A_{C} - \rm A_{S})} \cdot {S}_{\rm int}(\lambda)}{{S}_{\rm cou}(\lambda)}{C}_{\rm cou}(\lambda)},
\end{equation}
where $\rm E(\lambda)$ is the extinction curve of the observatory, $\rm A_{C}$ and $\rm A_{S}$ are the airmass of comparison star and standard star, respectively, ${S}_{\rm int}$ and ${S}_{\rm cou}$ are the intrinsic and observed spectrum of standard star, respectively. If the spectrum is not affected by the SOS, ${C}_{\rm fid}$ is equal to its intrinsic spectrum (${C}_{\rm int}$). Combining Equation~\eqref{eq2} and Equation~\eqref{eq3}, we can find that the contamination of SOS of object is a mixture of comparison star, standard star, and itself. Moreover, the atmospheric extinction varies with wavelength and weather, complicating the final SOS even further. Figure~\ref{Fig1} shows the fiducial spectra of comparison star that are produced by the spectra with (orange) and without (blue) SOS, respectively. The wavelengths in all figures are in observed-frame since this facilitates readers to estimate the observed influences of the SOS on other targets. There is a clear irregular shape in the red side of spectrum, and then propagate to the object during the relative flux calibration. We note that this effect only exists in term ${C}_{\rm int}$ of Equation~\eqref{eq2}. Thus, we can avoid the SOS of standard star if we already have ${C}_{\rm int}$, which can be easily obtained with only one observation on a clear night. After that, only ${O}_{\rm cou}$ and ${C}_{\rm cou}$ still contribute contamination of SOS in Equation~\eqref{eq2}. In principle, correcting for this effect requires knowledge of instrument information and exact weather conditions. This would be very complicated and almost impossible to achieve.

\begin{figure}[h]
  \begin{minipage}[t]{0.495\linewidth}
  \centering
   \includegraphics[width=60mm]{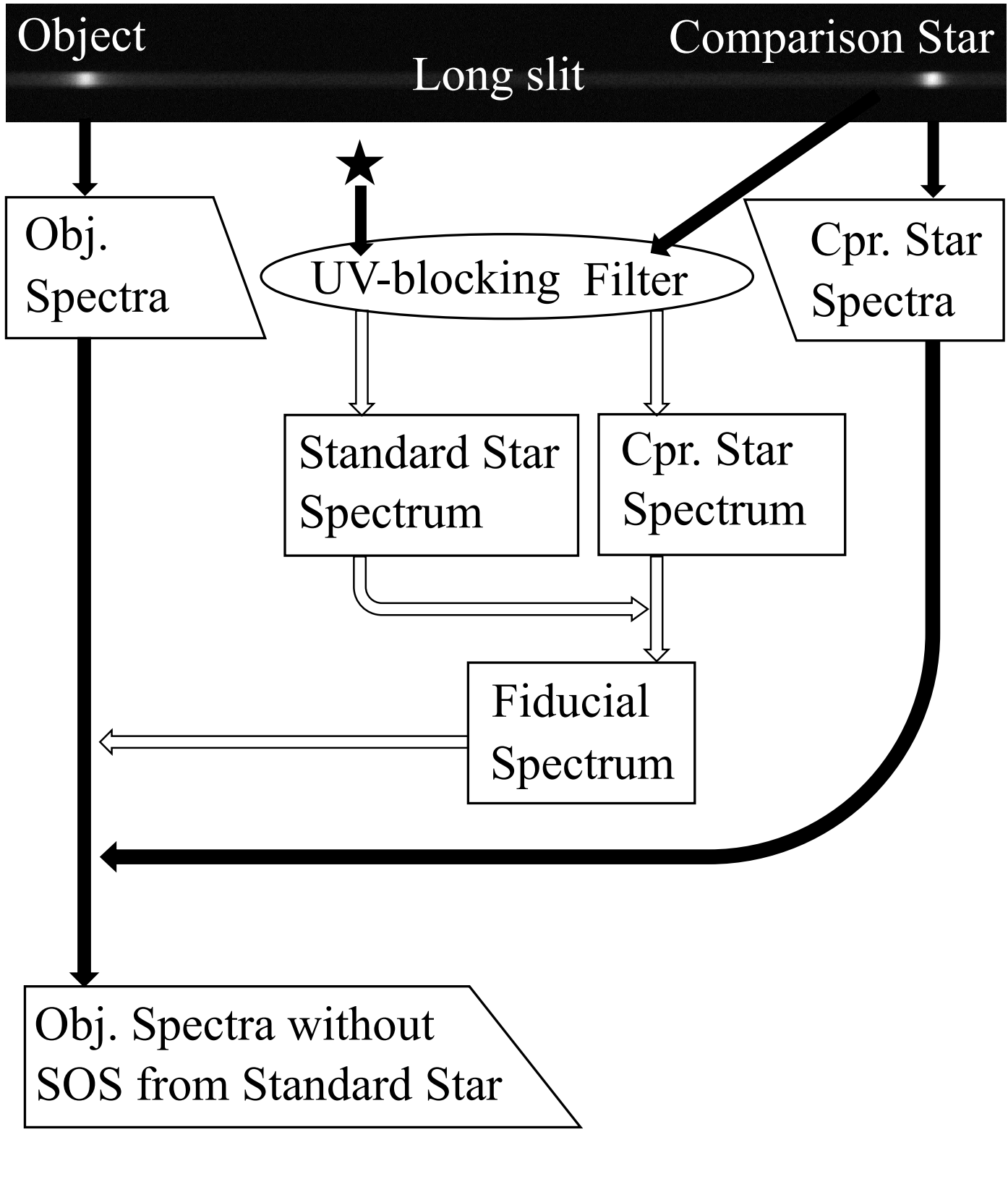}
  \end{minipage}%
  \begin{minipage}[t]{0.495\textwidth}
  \centering
   \includegraphics[width=60mm]{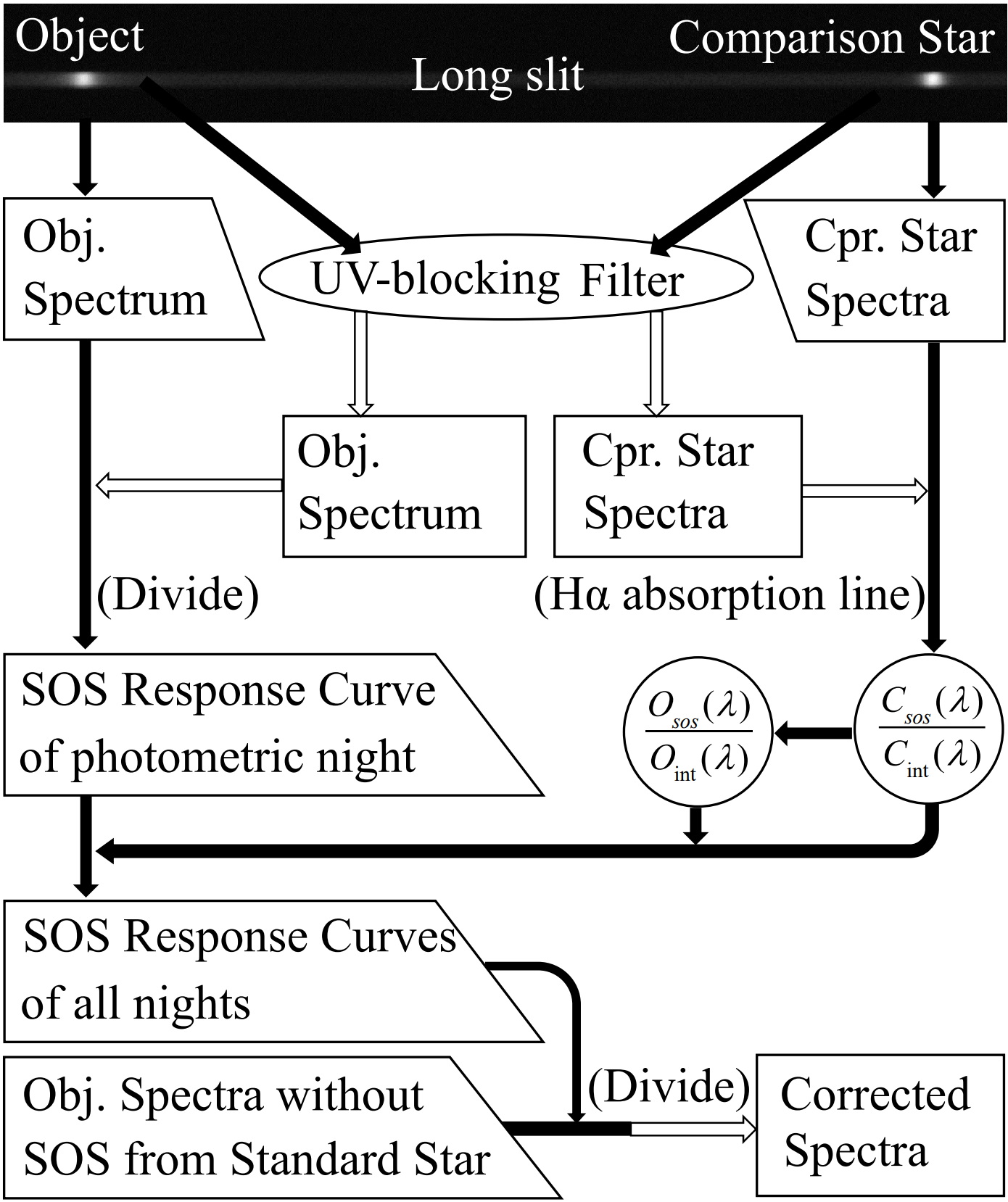}
  \end{minipage}%
  
  \caption{\footnotesize
   A flowchart for the method. Trapezoidal text boxes and black arrows represent for spectra contaminated by SOS. Oblong text boxes and white arrows represent for spectra without the contamination of SOS. The black five-pointed star is the standard star for absolute flux calibration.}
   \label{Fig2}
\end{figure}

A plausible approach is to assume that the spectral shape of AGN is constant during the monitoring period. Although, many observations show that there is a bluer-when-bright phenomenon in radio-loud AGNs \citep[e.g.,][]{Da21, Fa22, Ne22}, the assumption is still reasonable because (1) the RM projects usually focus on radio-quiet sources, and (2) the spectral index generally varies within a narrow range. Then, we can derive that the ratio of comparison star and object intrinsic spectra should be constant at any two wavelengths (i.e., $\lambda_{1}$ and $\lambda_{2}$)
\begin{equation}\label{eq4}
    \frac{{C}_{\rm int}(\lambda)}{{O}_{\rm int}(\lambda)} \propto \frac{{C}_{\rm int}(2\lambda)}{{O}_{\rm int}(2\lambda)}.
\end{equation}
Besides, the SOS efficiencies ($\eta$) of comparison star and object should be consistent with each other due to the identical weather conditions and instrument, i.e., there SOS can be expressed as
\begin{equation}\label{eq5}
    {C}_{\rm sos}(2\lambda) = \eta {C}_{\rm int}(\lambda),
\end{equation}
\begin{equation}\label{eq6}
    {O}_{\rm sos}(2\lambda) = \eta {O}_{\rm int}(\lambda).
\end{equation}
Combining Equation~\eqref{eq4}, Equation~\eqref{eq5}, and Equation~\eqref{eq6}, we can obtain that the ratio of the comparison star SOS to the intrinsic flux density should be proportional to that of the object
\begin{equation}\label{eq7}
    \frac{{C}_{\rm sos}(\lambda)}{{C}_{\rm int}(\lambda)} \propto \frac{{O}_{\rm sos}(\lambda)}{{O}_{\rm int}(\lambda)}.
\end{equation}
This means that the SOS of object can be derived from the comparison star. To measure the SOS of comparison star in each epoch, we decompose the spectrum into first- and second-order components (i.e., ${C}_{\rm cou1}$ and ${C}_{\rm cou2}$) and calculate a sensitivity function using two windows of line-free regions around H$\alpha$ absorption line
\begin{equation}\label{eq8}
    sens = \frac{{C}_{\rm cou1}(\lambda) + {C}_{\rm cou2}(\lambda)}{{C}_{\rm int}(\lambda)}.
\end{equation}
In principle, if the SOS is smooth, then the flux of continuum-subtracted absorption line should not be contaminated by SOS. However, after rewriting the two components to $1 + {{C}_{\rm sos}(\lambda)}/{{C}_{\rm int}(\lambda)}$ by dividing ${C}_{\rm cou1}$, the $sens$ will introduce a fraction of SOS to the calibrated H$\alpha$ absorption flux
\begin{equation}\label{eq9}
    F_{\rm cal} = \frac{F_{\rm cou}}{sens}
    =\frac{F_{\rm cou}}{C_{\rm cou1}(\lambda)}
    \frac{C_{\rm int}(\lambda)}{ 1 + \frac{{C}_{\rm cou2}(\lambda)}{{C}_{\rm cou1}(\lambda)} }   
    = \frac{F_{\rm int}}{1 + \frac{{C}_{\rm sos}(\lambda)}{{C}_{\rm int}(\lambda)}},
\end{equation}
where $F_{\rm int}$ is the intrinsic flux of H$\alpha$ absorption line which can be measured from a non-SOS spectrum. 
Note that this pertains only to the SOS under the assumption of a smooth SOS. ${C}_{\rm sos}(\lambda)/{{C}_{\rm int}(\lambda)}$ represents the ratio of the superimposed SOS of the comparison star, from which we can derive ${{O}_{\rm sos}(\lambda)}/{{O}_{\rm int}(\lambda)}$ and consequently, obtain the corrected spectra of the scientific object. Figure~\ref{Fig2} is a flowchart to illustrate our correction scheme. Although Equation~\eqref{eq4} and Equation~\eqref{eq7} suggest that the relative change of SOS at different wavelengths should be consistent with each other, this still requires observations to confirm, and we also need to examine the shape of the SOS.

In general, to apply the method to existing time-domain observational data, we need to re-observe the uncontaminated standard, comparison and object on a photometric night. This will yield a reliable fiducial spectrum of the comparison star and one response function that represents the SOS contamination of the object/comparison star pair for each day, taking advantage of the strategy that observing a nearby comparison star simultaneously. 

\section{Application to NGC~5548}

\label{sec3}

The Yunnan Faint Object Spectrograph and Camera (YFOSC), equipped with a series of grisms, long slits and filters, 
is a versatile instrument for spectroscopy and photometry mounted on the 2.4 m telescope. 
Before 2018, the ultraviolet-blocking (UV-blocking) filter that was usually used to block the SOS overlapping on the scientific spectrum was not equipped in YFOSC, 
so the spectral regions where contaminated by SOS in archive spectra (here after contaminated spectra) were not well considered. 
For example, NGC~5548 was well monitored by a long-term RM campaign utilizing the Lijiang 2.4~m telescope to probe the BLR evolution. \cite{Lu+etal+2022} performed a five-season observation and published the RM results of broad H$\gamma$, H$\beta$ and Helium line, while the broad H$\alpha$ line was not considered because the H$\alpha$ region was contaminated by SOS. 
In this section, we apply above method to eliminate the overlapped SOS from NGC~5548 data observed in 2019. So that we can obtain the intrinsic spectrum, then compare the resulting light curve of broad H$\alpha$ line before and after eliminating SOS, to estimate the impact of SOS on the time series. 
\subsection{Observation and Data Reduction}
As described in Section 2, constructing the uncontaminated fiducial spectrum is the core step for eliminating the overlapping SOS from the scientific data. 
During the RM spectroscopic monitoring of NGC~5548 in 2019, we selected a photometric night (JD=2458465), and performed a spectroscopic observation adding a UV-blocking filter to obtain the UV cut-off spectra of NGC 5548 and its comparison star. 
Because the filter cuts off at approximately 4150$\rm{\mathring{A}}$, the spectra at wavelength below 8300$\rm{\mathring{A}}$ were not affected by SOS. Then a spectrophotometric standard star (G191-B2B) was also observed with the UV-blocking filter in the nearby sky, 
which is used to generate the UV-blocking fiducial spectrum. 
In addition, we carried out six spectroscopic observations after RM observation in the six nights, and obtained extra UV cut-off spectra of NGC~5548 and its comparison star, which are used to check the validity of our designed method. 

Following the work of \cite{Lu+etal+2022}, we reduced the the UV cut-off spectra using standard IRAF version 2.16 routines. The primary routines include bias subtraction, flat-field correction, wavelength calibration, and cosmic-ray elimination. Standard neon and helium lamps were used for wavelength calibration. The one-dimensional was extracted with aperture of 5.66$^{\prime\prime}$. 
Varying seeing and mis-centering usually leaded to wavelength shifts and we used the [O {\sc iii}] $\rm{\lambda5007}$ line as the wavelength reference with the interpolated cross-correlation function (ICCF).

\begin{figure}
   \centering
   \includegraphics[width=\textwidth, angle=0]{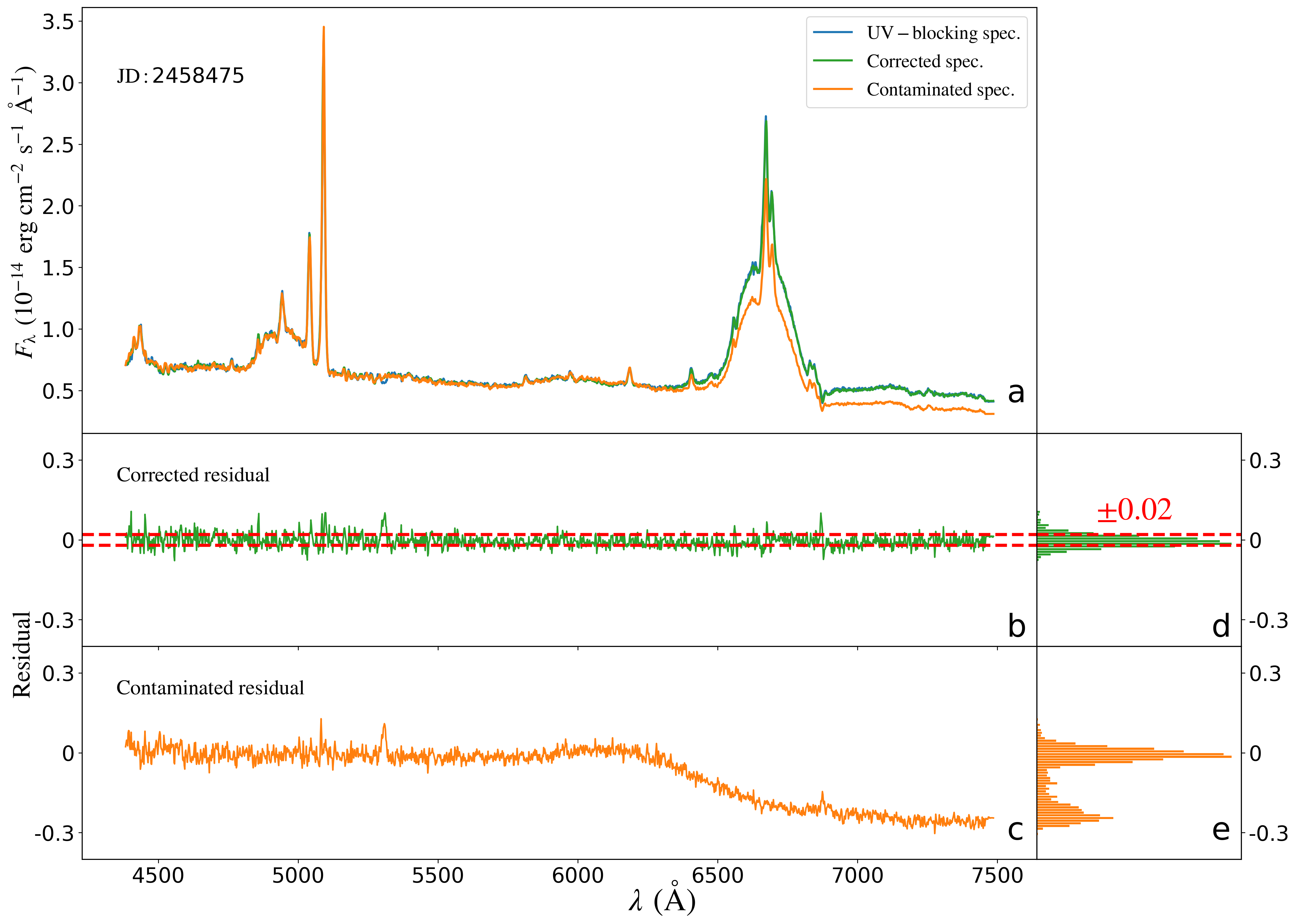}
   \caption{\footnotesize
    Results of JD 2458475. a. Spectra of NGC 5548 from the UV-blocking observation (blue), contaminated observation (orange) and the correction method (green). 
    b. Residual of contaminated spectra relative to UV-blocking spectra. c. Residual of corrected spectra relative to UV-blocking spectra. d. Hist of the residual of contaminated spectra. e. Hist of the residual of corrected spectra. The red dash lines are 2\% and -2\%. Note the blue line is the result from the UV-blocking data, which almost perfectly coincides with the green line. 
    }
   \label{Fig3}
\end{figure}

\subsection{Eliminating the contamination of SOS}
From the UV-blocking spectroscopic observation in the photometric night, we generated the UV-blocking fiducial spectrum through the calibration process of spectral flux using the UV cut-off spectra of standard star. Meanwhile, we obtained a contaminated fiducial spectrum from the contaminated spectra observed in 2019 RM campaign. The term ``contaminated'' means that the spectrum are overlapped by SOS because of the lack of the UV-blocking filter. Figure~\ref{Fig1} clearly shows that the spectrum at the regions of wavelength larger than 6300 angstrom are contaminated by SOS. This means that if the scientific spectrum is calibrated by the contaminated fiducial spectrum, we cannot obtain the intrinsic spectrum in the regions overlapped by SOS. In this section, we use the method described in Section~\ref{sec2} to construct the intrinsic spectrum for the contaminated spectra of NGC~5548 observed in observing season of 2019 (\citep{Lu+etal+2022}).

\begin{figure}
   \centering
   \includegraphics[width=\textwidth, angle=0]{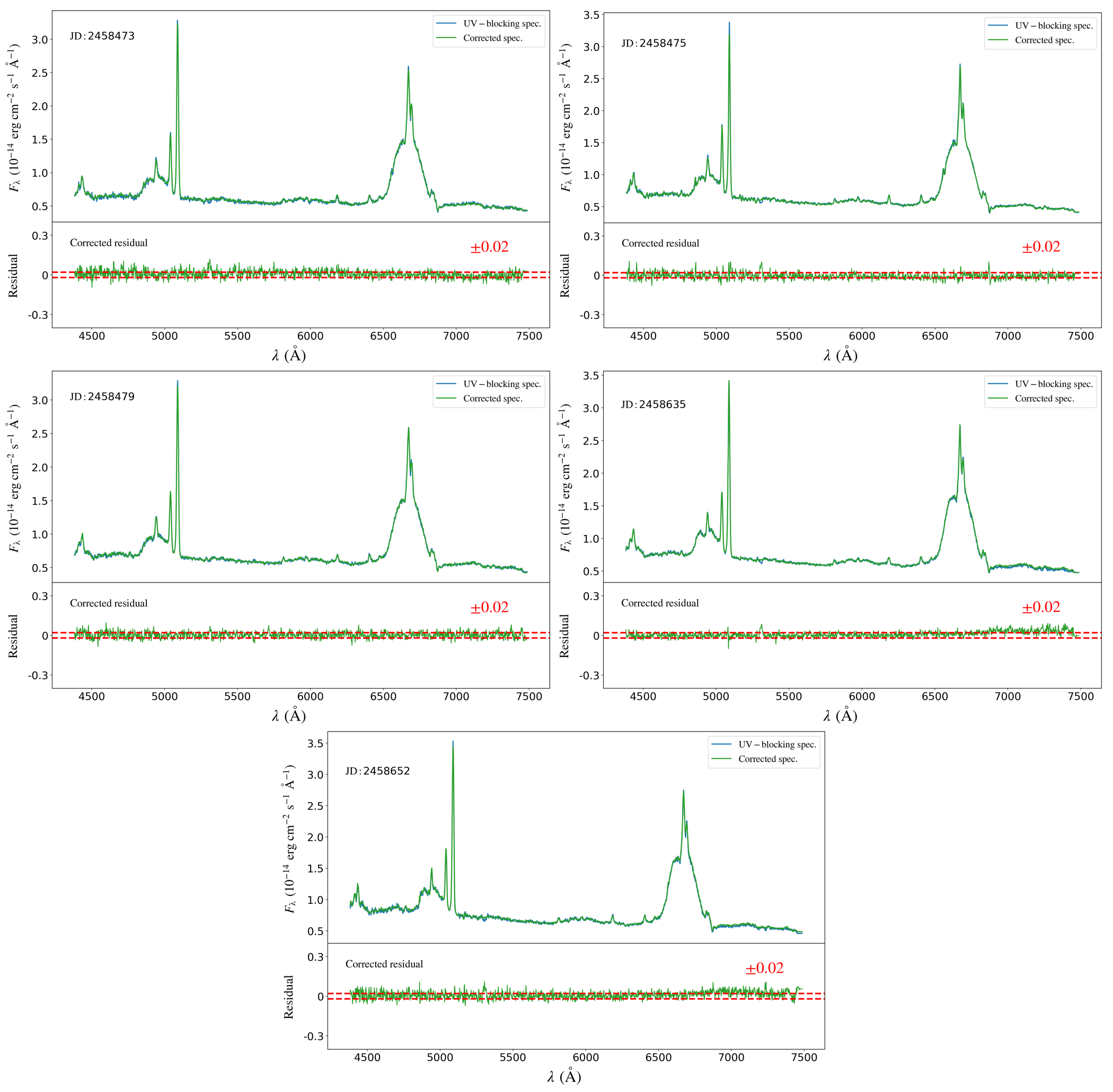}
   \caption{\footnotesize
   Results of the 5 days. The green lines are corrected spectra, and the blue lines are contaminated spectra. }
   \label{Fig4}
\end{figure}

Following the method described in Section~\ref{sec2}, we first obtained the contaminated sensitivity function by comparing the observed spectrum of the comparison star in each exposure to the UV-blocking fiducial spectrum. Then this sensitivity function was applied to calibrate the contaminated spectrum of NGC 5548 observed during the nights when we conducted the UV-blocking spectroscopic observations. At the same time, we generated a UV cut-off sensitivity function and used it to calibrate the UV cut-off spectrum of NGC~5548, resulting in the UV-blocking spectrum. This step helped us to eliminate the contamination from the standard star, and the UV-blocking spectrum contains no SOS at all. To obtain $F_{\rm{cal}}$ in Equation~\eqref{eq9}, we used a compound model consisting of a linear and a Gaussian model to fit the spectra of the comparison star near the H$\alpha$ absorption band. From Equation~\eqref{eq7} to \eqref{eq9}, we derived $O_{\rm sos}$ and ${C}_{\rm sos}$ and computed one response factor for each night which represents the ratio of the night to the photometric night in terms of the response functions of SOS. The response function of the contamination from the object/comparison pair in photometric night was generated by the two calibrated spectra obtained in the last step. Then, together with the factor, the corrected spectra of the NGC~5548 in the six days were finally derived. 

Panel (a) of Figure~\ref{Fig3} shows the spectra of NGC~5548 for one day, with the UV-blocking and corrected spectrum shown in blue and green, respectively. The contaminated spectrum, which were calibrated ignoring SOS, is shown in orange. Panel (b) gives the $\sim$2\% deviations between the corrected spectrum and the UV-blocking spectrum, which demonstrates that the contaminated spectrum is well corrected by our designed method. Results of the five days are presented in Figure~\ref{Fig4}, where the deviations are also negligible (there were clouds in the Julian date 2458635). These results indicate that our method of processing contaminated data was successful in producing results that were nearly identical to the UV-blocking spectra. Therefore, we proceeded to apply this correction method to the 81 contaminated spectra. 

\begin{figure}
   \centering
   \includegraphics[width=\textwidth, angle=0]{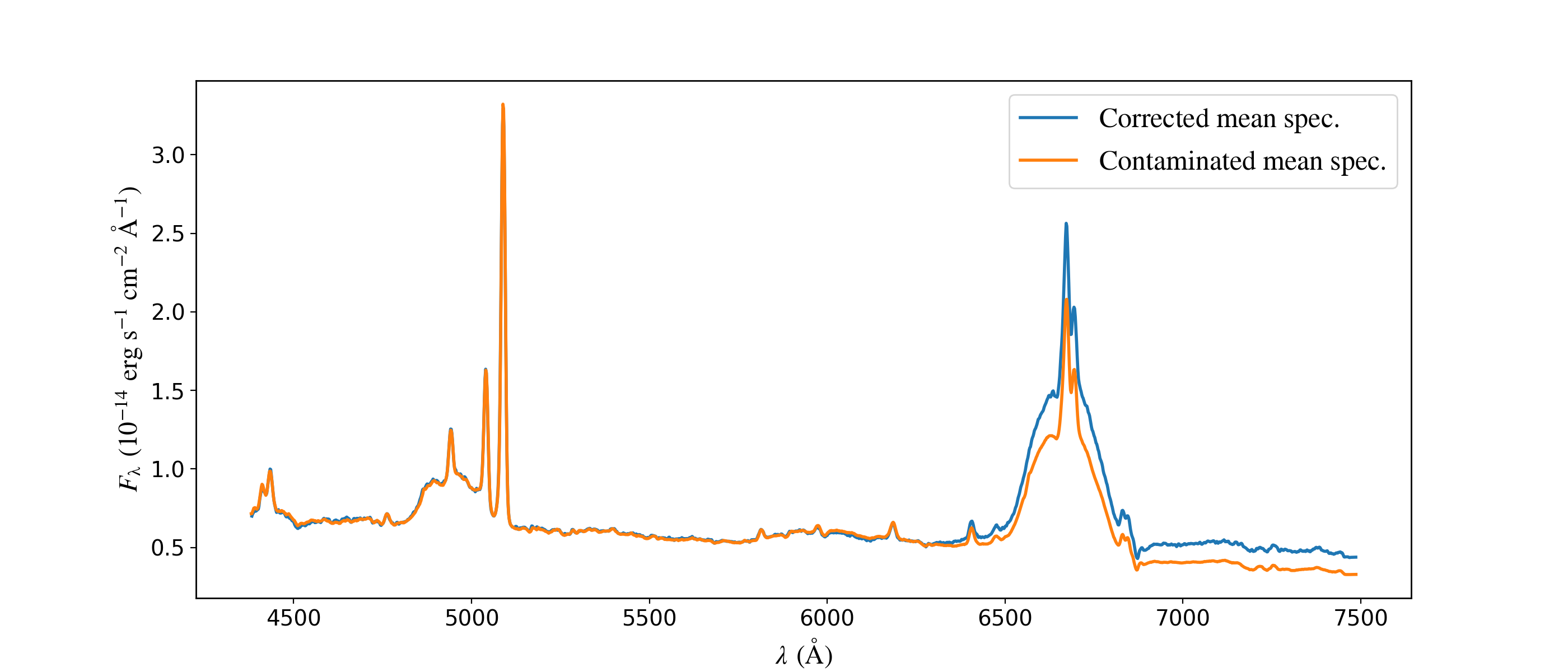}
   \caption{\footnotesize
   Mean spectra of NGC~5548. The orange line is the mean spectrum of the contaminated spectra and the blue line is of the corrected spectra.}
   \label{Fig5}
\end{figure}

\subsection{Comparison}
We generated the mean contaminated and corrected spectrum of the 81 spectra and displays them in Figure~\ref{Fig5}. Panel (a) of Figure~\ref{Fig3} also presented the contaminated spectrum and its deviation from the UV-blocking spectrum shown in Panel (c). These figures revealed that the impact of SOS on the absolute spectral flux can be up to 30\%. 

We generated $\rm{H\alpha}$ light curves using both the contaminated and corrected spectra to assess the impact of SOS. Light curves of the continuum (rest-frame 5100 $\rm{\mathring{A}}$) and the $\rm{H\beta}$ emission line are also presented as references because they are not affected by SOS. However, the contaminated $\rm{H\alpha}$ line does not exhibit the typical spectral shape. Therefore, we measured the light curve using integration rather than spectral decomposition. This approach was also employed for the other two light curves for comparison.

\begin{figure}
   \centering
   \includegraphics[width=\textwidth, angle=0]{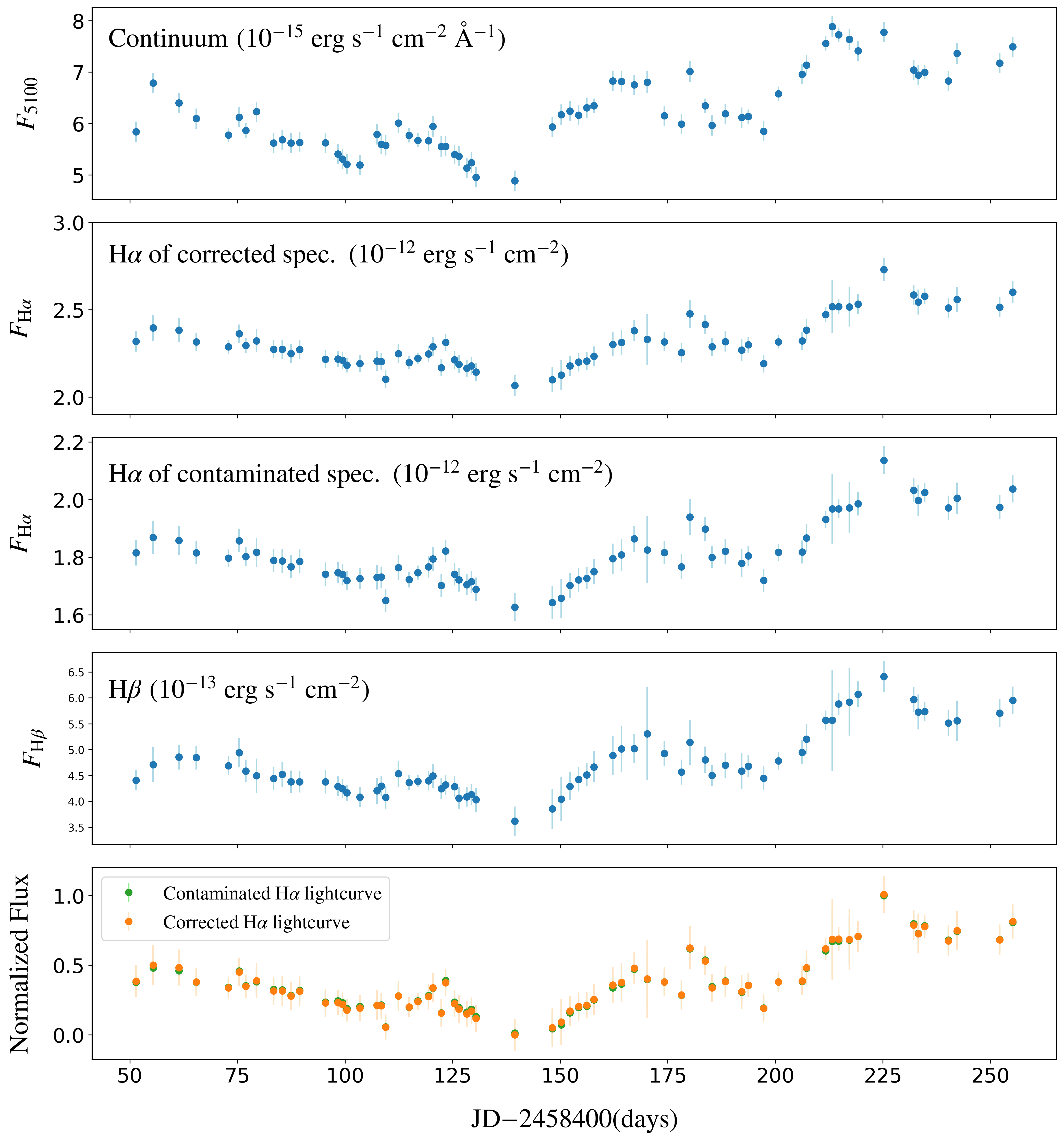} 
   \caption{\footnotesize
   Light Curves of NGC 5548. From top to bottom: 1. $F_{5100}$ light curve. 2. $\rm{H\alpha}$ light curve using corrected spectra. 3. $\rm{H\alpha}$ light curve using contaminated spectra. 4. $\rm{H\beta}$ light curve. 5. Normalized $\rm{H\alpha}$ light curves. }
   \label{Fig6}
\end{figure}

To measure the continuum light curve, we calculated the median value between 5090 and 5110 $\rm{\mathring{A}}$. For the $\rm{H\alpha}$ emission line, we inspected the mean spectra and confirmed the red, blue, and integral window as 6180-6250 $\rm{\mathring{A}}$, 6880-6850 $\rm{\mathring{A}}$ and 6400-6700 $\rm{\mathring{A}}$, respectively. We used the red and blue windows for continuum fitting, and then integrated flux of the emission line in the integral window after subtracting the fitted continuum. As for $\rm{H\beta}$, the windows were 4500-4520 $\rm{\mathring{A}}$, 5090-5100 $\rm{\mathring{A}}$ and 4700-4920 $\rm{\mathring{A}}$, respectively. The light curves were presented in Figure~\ref{Fig6}. It is evident from the figure that the two $\rm{H\alpha}$ light curves exhibit a similar shape, indicating that the impact on time-lag is negligible. 

\section{Discussion}
\label{sec4}

In most RM campaigns that do not specifically focus on the $\rm{H\alpha}$ line, there are often overlaps between the $\rm{H\alpha}$ line and the SOS of the UV/blue band. The practical impact of SOS depends on the relative blueness of the standard star and the object/comparison star pair. Based on our analysis of NGC~5548, we can deduce that the standard star (G191-B2B) was much bluer than the object/comparison star pair, and NGC 5548 was slightly bluer than J1417. As a result, the flux of contaminated spectra with ignored SOS in the flux calibration process was significantly lower than that of the corrected spectra around the $\rm{H\alpha}$ band. This was primarily due to the division of a larger sensitivity function generated from the comparison star, and the SOS-effect of the object/comparison star pair tends to reduce the effect caused by the contaminated fiducial spectrum. The overall impact is about 30\% and this indicates that Grism14, which was used for monitoring NGC~5548 and most other objects of the LJT (hereafter used to refer to Lijiang 2.4 m telescope) RM campaign, has a significant impact on absolute flux calibration, as shown in Figure~\ref{Fig5}. As mentioned above, the factor represents for the systematic bias is mainly caused by G191-B2B and this factor depends on the spectral slopes of the object, comparison star, and standard star.

As demonstrated in Figure~\ref{Fig6}, the light curve shows less susceptibility to the SOS, except for the integrated absolute flux. First, this could be attributed to the higher sensitivity of the light curves towards relative variations in flux, as compared to nearly proportional changes in absolute flux that have been contaminated by almost the same ratio everyday in $\rm{H\alpha}$ band. Secondly, partial SOS-effect was eliminated by subtracting the linear-fitting continuum before integrating the emission-line flux. However, the broadness of the $\rm{H\alpha}$ integral interval renders the absolute flux highly sensitive to changes in spectral shape caused by SOS in contrast to that of the $\rm{H\alpha}$ absorption line of the comparison star, and this impact is relatively small compared with the proportional changes caused by SOS. 

The contamination of the object/comparison star pair changes in different observation conditions, leading to a slight variation on the contaminated $\rm{H\alpha}$ light curves. This effect will cause the light curve contaminated by SOS to closely resemble the continuum light curve, as the origin of SOS primarily from the continuum in the UV/blue band. The lack of spectral decomposition and host galaxy contamination elimination, which is for the sake of consistency with the contaminated results, makes the continuum light curve not as good as the previous work (\citep{Lu+etal+2022}). But it is sufficient to observe the trend of changes in the H$\alpha$ light curve caused by the contamination of SOS. Multi-line RM observations indicate that the $\rm{H\alpha}$ line is emitted from much larger region than $\rm{H\beta}$ line \citep{Be10a, Fe21b}, and its light curve should be smoother. However, as shown in Figure~\ref{Fig6}, the opposite was observed with SOS overlapping. 
The reason for this may be the different presence of components within the integral range. It is evident that the correction consistently decreases the normalized flux when the light curve is trending downward, and conversely, increases it when trending upward. This implies that the correction method slightly shifts the light curve to the left.

The basically proportional variation caused by SOS on absolute flux of H$\alpha$ line is approximately 30\%. Further more, the emission line width and spectral shape significantly impact RM measurements, such as the determination of SMBH masses and the velocity-resolved time-lags. We can see that our method corrected the line profile significantly. Correcting for the spectral shape enables exploration of the structure of the H$\alpha$-emitting BLR by velocity-resolved time-lags. Therefore, further research on $\rm{H\alpha}$ RM measurements using this correction method will be conducted in the next paper.

The corrected spectra only has a deviation of about 2\% compared to the UV-blocking spectra obtained from an order-blocking filter. We anticipate that we can have even less deviation by correcting the telluric absorption of the comparison star using the method proposed by \cite{Lu+etal+2021}. It is worth mentioning that when we apply this method to specific archival data, a minority of AGNs might undergo a significant variation in spectral shape rendering the basic assumption of the method, namely, that the spectral shape remains constant, no longer valid. Therefore, we suggest quantifying the variation if there have been a long elapsed time since the archival observations. By applying this method to contaminated data, we can carry out expensive RM experiments without consuming too much telescope time, as theoretically only one day of optimal weather conditions is required for each object. 
We plan to apply this method to archival RM data, especially that of the LJT, which has a potential to increase the current $\rm{H\alpha}$ RM measurements. 

\section{Summary}
\label{sec5}

In this paper, we present a method to correct the contamination of SOS using a simultaneously observed comparison star. To derive corrected spectra from SOS-contaminated data, we implemented a two-step procedure. First, we eliminate the SOS of the spectrophotometric standard star by generating an uncontaminated fiducial spectrum. This fiducial spectrum is then utilized to calculate the sensitivity function of object and comparison star. Second, we eliminate the SOS of the object/comparison pair using the absorption feature of the H$\alpha$ line of comparison star.

We tested this method on NGC~5548 by obtaining additional seven spectra with a UV-blocking filter. 
An uncontaminated fiducial spectrum and a response function of SOS were derived from observations made in a photometric night. The rest six UV-blocking spectroscopic observations were used to obtain uncontaminated UV-blocking spectra, along with the corrected ones, to verify the validity of this correction method, resulting in an approximately 2\% deviation. 

This method was applied to RM spectra obtained during the 2019 observation season, resulting in 81 corrected spectra, together with 81 contaminated spectra which were calibrated ignoring the contamination of SOS. 
The corrected spectra were compared to the contaminated spectra to assess the influence of SOS. The major impact of SOS is manifested in the following aspects. 

\begin{enumerate}
\item For absolute flux near H$\alpha$, the contaminated spectra is about 30\% lower than the UV-blocking spectra. 
\item The spectral shape changes significantly making line width and velocity-resolved time-lag measurements unreliable for further investigation if we directly use the contaminated spectra. 
\item The contaminated H$\alpha$ light curve is less smooth than H$\beta$ and the timing of changes in the corrected H$\alpha$ light curve has advanced very slightly.
\end{enumerate}

Our application of the method can successfully correct the spectral shape, as evidenced by comparing the UV-blocking results. 

\begin{acknowledgements}
We express our sincere gratitude to the reviewer for the meticulous reading and constructive comments. 
This work is funded by the National Key R\&D Program of China with No. 2021YFA1600404, the National Natural Science Foundation of China (NSFC; grant Nos. 11991051, 12303022, 12373018, 12203096, 12103041, and 12073068), Yunnan Fundamental Research Projects (grant Nos. 202301AT070339 and 202301AT070358), Yunnan Postdoctoral Foundation Funding Project, the Yunnan Province Foundation (202001AT070069), the Youth Innovation Promotion Association of the Chinese Academy of Sciences (2022058), the Topnotch Young Talents Program of Yunnan Province, Special Research Assistant Funding Project of Chinese Academy of Sciences, and the science research grants from the China Manned Space Project with No. CMS-CSST-2021-A06. We acknowledges the staff of the Lijiang Station of the Yunnan Observatory for their enthusiasm and for operating the LJT so efficiently and professionally. Funding for the telescope has been provided by the Chinese Academy of Sciences and the People's Government of Yunnan Province. 
\end{acknowledgements}

\label{lastpage}

\end{document}